\documentclass[preprint,showpacs,preprintnumbers,amsmath,amssymb,endfloats]{revtex4}
\usepackage{graphicx}
\usepackage{dcolumn}
\usepackage{bm}
\usepackage{amssymb}
\usepackage{amsmath}
\usepackage{latexsym}
\usepackage{longtable}
\usepackage{color}

\begin{document}

\title{Applicability of the absence of equilibrium in quantum  system fully coupled to several fermionic and bosonic heat baths}

\author{V.V. Sargsyan$^{1}$, A.A. Hovhannisyan$^{1,2,3}$,     G.G. Adamian$^{1}$,  N.V. Antonenko$^{1,4}$, and D. Lacroix$^{5}$ }
\affiliation{
$^{1}$Joint Institute for Nuclear Research, 141980 Dubna, Russia\\
$^{2}$Institute of Applied Problems of Physics,  0014 Yerevan, Armenia\\
$^{3}$Quantum Computing Laboratory,   1142 Norakert, Armenia \\
$^{4}$Tomsk Polytechnic University, 634050 Tomsk, Russia\\
$^{5}$Universit{\'e}
Paris-Saclay, CNRS/IN2P3, IJCLab, 91405 F-91406 Orsay Cedex, France
}
\affiliation{}
\date{\today}

\begin{abstract}
The time evolution of occupation number is studied for fermionic or bosonic oscillator
linearly fully coupled  to  several fermionic and bosonic heat baths.
The influence of characteristics of thermal reservoirs of different statistics on the
non-stationary population probability is analyzed  at large times.
Applications of the absence of equilibrium in such systems for  creating a dynamic (nonstationary) memory storage  are  discussed.
%
%
\end{abstract}
\maketitle

\section{Introduction}
The quantum  systems are never completely isolated and interact with
 large number of degrees of freedom of  surrounding  environment.
The coupling of a quantum system to a heat bath usually induces its evolution towards an asymptotic
 equilibrium imposed by the complexity of the heat bath(s). In practice,
a quantum system is often coupled to a few reservoirs \cite{Stefanescu,18,19,17,16,20,FFF,PRE18,wen,PRA2020}.
In Refs. \cite{PhysicaA2019,PRE2020},
it have been illustrated that the system linearly fully coupled  to several baths of different statistics (fermionic and bosonic)
might never reach a stationary asymptotic limit.
This absence of equilibrium at large times  can be used in some  applications,
for example, in the communication lines,   quantum computers, and other modern quantum devices.
So,
 in the present paper  we study   the time evolution of occupation numbers of
 fermionic (two-level system) and bosonic oscillators embedded in the fermionic and bosonic heat baths.
 A system  fully coupled  to two heat baths with the same or different quantum natures
is  described here using the non-Markovian master-equation and quantum Langevin approaches \cite{PRE2020},
and taking into consideration the Ohmic dissipation with Lorenzian cutoffs \cite{M1,Haake,Dodonov,Isar,Armen}.
The full coupling  contains the resonant [the rotating wave approximation] and non-resonant terms \cite{Armen}.
The  environmental effects  on a quantum
system could keep this system in certain state or provide it some specific properties.

\section{Model}

\subsection{Hamiltonian}
The Hamiltonian of the total system  (the quantum system plus  several heat baths "$\lambda$", $\lambda=1,\cdots,N_{b}$)  is
written as \cite{PRE2020}
\begin{equation}
\label{ham}
H=H_c+\sum_{\lambda=1}^{N_b} H_{\lambda} +  \sum_{\lambda=1}^{N_b} H_{c,\lambda},
\end{equation}
where
\begin{equation}
H_c=\hbar \omega a^{\dagger} a
\end{equation}
is the Hamiltonian of the  isolated  system being either fermionic (two-level system)
or bosonic oscillator with   frequency $\omega$,
\begin{equation}
\label{eq:baths}
H_{ \lambda}= \sum_i \hbar \omega_{\lambda,i} c^{\dagger}_{\lambda,i} c_{\lambda,i}  \nonumber
\end{equation}
are the Hamiltonians of the thermal baths.  When we write down the creation/annihilation operators $a^+$/$a$  ($c^{\dagger}_{\lambda,i}/c_{\lambda,i}$)
we mean the creation/annihilation  operators of transition with the
corresponding energy $\hbar \omega$ ($\hbar \omega_{\lambda,i}$).
So, each fermionic transition operator $a^+$   or  $c^{\dagger}_{\lambda,i}$  is the product of
operators of creation and annihilation of a fermion in the excited and ground states, respectively.
There
is only the conversion of excitation quanta from fermionic system to the
bosonic ones or vice versa in our formalism.
The value of $N_{b}$ is the number of   heat baths.
Each heat bath "$\lambda$" is  modeled by the assembly of  independent fermionic or bosonic oscillators
labelled in both cases by "$i$" with frequencies  $\omega_{\lambda,i}$.
For the FC coupling between the  system and heat baths,
the interaction Hamiltonians $H_{c,\lambda}$ are
\begin{equation}
\label{eq:int1}
H_{c, \lambda}= \sum_i \alpha_{\lambda,i} ( a^{\dagger}+a )(c^{\dagger}_{\lambda,i}+c_{\lambda,i}).
\end{equation}
The  real constants $\alpha_{\lambda,i}$ determine the coupling strengths.
The  interaction Hamiltonian (\ref{eq:int1}) is  linear in the system and baths operators.
It  has important consequences on the dynamics of the  system by
altering the effective collective potential and by allowing energy to be exchanged with the thermal reservoirs,
thereby, allowing the  system to attain   some equilibrium with the heat baths.

Here, the  system and heat baths have the fermionic or bosonic statistics.
So, the creation  and annihilation   operators of  the system and heat baths
satisfy the  commutation or anti-commutation relations:
\begin{equation}
\label{eq:perm}
\begin{split}
&aa^{\dagger}-\varepsilon_{a} a^{\dagger}a=1,\quad a^{\dagger}a^{\dagger}-\varepsilon_{a} a^{\dagger}a^{\dagger}=aa-\varepsilon_{a} aa=0,\\
&c_{\lambda,i}c^{\dagger}_{\lambda,i}-\varepsilon_\lambda  c^{\dagger}_{\lambda,i}c_{\lambda,i}=1,
\quad c^{\dagger}_{\lambda,i}c^{\dagger}_{\lambda,i}-\varepsilon_\lambda c^{\dagger}_{\lambda,i}c^{\dagger}_{\lambda,i}=c_{\lambda,i}c_{\lambda,i}-\varepsilon_\lambda  c_{\lambda,i}c_{\lambda,i}=0,
\end{split}
\end{equation}
where  $\varepsilon_{a} $ and $\varepsilon_\lambda $ are equal to 1 (-1)
for the bosonic (fermionic) system and bosonic (fermionic) heat baths, respectively.

\subsection{Master-equation for occupation number of quantum system}

Employing the Hamiltonian (\ref{ham}) for the fermionic  and bosonic  systems, we deduce the  equations of motion
for the occupation number
\begin{eqnarray}
\frac{d a^\dagger(t)a(t)}{dt}&=&
\frac{i}{\hbar}\sum_{\lambda,i}\alpha_{\lambda,i}[a(t)-a^{\dag}(t)][c^{\dagger}_{\lambda,i}(t)+c_{\lambda,i}(t)]\nonumber \\
&=&
\frac{i}{\hbar}\sum_{\lambda,i}\alpha_{\lambda,i}[c^{\dagger}_{\lambda,i}(t)a(t)-a^{\dag}(t)c_{\lambda,i}(t)+
a(t)c_{\lambda,i}(t)-a^{\dag}(t)c^{\dagger}_{\lambda,i}(t)].
\label{master-eq}
\end{eqnarray}
For the operators
$c^{\dagger}_{\lambda,i}(t)a(t)$ and
$a(t)c_{\lambda,i}$
in Eq. (\ref{master-eq}),
we derive the following equations
\begin{eqnarray}
\frac{d c^{\dagger}_{\lambda,i} a}{dt}  & = &  i(\omega_{\lambda,i}  - \omega)  c^{\dagger}_{\lambda,i} a + \frac{i}{\hbar}  \alpha_{\lambda,i}[a^\dagger a+a a][1-(1-\varepsilon_\lambda)c^{\dagger}_{\lambda,i} c_{\lambda,i}] \nonumber \\
&-& \frac{i}{\hbar}\sum_{\lambda',i'}\alpha_{\lambda',i'}[c^{\dagger}_{\lambda,i}c^{\dagger}_{\lambda',i'}+c^{\dagger}_{\lambda,i}c_{\lambda',i'}]
[1-(1-\varepsilon_{a})a^+a],
\label{eq:ur3} \\
\frac{dac_{\lambda,i}}{dt}  & = & -i(\omega_{\lambda,i}  + \omega)     a c_{\lambda,i} - \frac{i}{\hbar}\alpha_{\lambda,i}[a^\dagger a+a a][1-(1-\varepsilon_\lambda)c^{\dagger}_{\lambda,i}c_{\lambda,i}] \nonumber \\
&-& \frac{i}{\hbar}\sum_{\lambda',i'}\alpha_{\lambda',i'}[c_{\lambda,i}c^{\dagger}_{\lambda',i'}+c_{\lambda,i}c_{\lambda',i'}][1-(1-\varepsilon_{a})a^+a].
\label{eq:ur4}
\end{eqnarray}
Substituting the formal solutions
\begin{eqnarray}
&&c^{\dagger}_{\lambda,i}(t)a(t)= e^{i(\omega_{\lambda,i}  - \omega) t}c^{\dagger}_{\lambda,i}(0)a(0) \nonumber \\
&+& \frac{i}{\hbar }\alpha_{\lambda,i}\int^t_0d\tau e^{i(\omega_{\lambda,i}  - \omega) [t-\tau]}[a^\dagger(\tau) a(\tau)+a(\tau) a(\tau)][1-(1-\varepsilon_\lambda)c^{\dagger}_{\lambda,i}(\tau) c_{\lambda,i}(\tau)] \nonumber \\
&-& \frac{i}{\hbar }\sum_{\lambda',i'}\alpha_{\lambda',i'}\int^t_0d\tau e^{i(\omega_{\lambda,i}  - \omega) [t-\tau]}[c^{\dagger}_{\lambda,i}(\tau)c^{\dagger}_{\lambda',i'}(\tau)+c^{\dagger}_{\lambda,i}(\tau)c_{\lambda',i'}(\tau)][1-(1-\varepsilon_{a})a^+(\tau)a(\tau)],
\nonumber  \\
&&a(t)c_{\lambda,i}(t)= e^{-i(\omega_{\lambda,i}  + \omega) t}a(0)c_{\lambda,i}(0) \nonumber \\
&-& \frac{i}{\hbar } \alpha_{\lambda,i}\int^t_0d\tau e^{-i(\omega_{\lambda,i}  + \omega)  [t-\tau]}[a^\dagger(\tau) a(\tau)+a(\tau) a(\tau)][1-(1-\varepsilon_\lambda)c^{\dagger}_{\lambda,i}(\tau)c_{\lambda,i}(\tau)]
\nonumber \\
&-& \frac{i}{\hbar}\sum_{\lambda',i'}\alpha_{\lambda',i'}\int^t_0d\tau e^{-i(\omega_{\lambda,i}  + \omega)  [t-\tau]}[c_{\lambda,i}(\tau)c^{\dagger}_{\lambda',i'}(\tau)+c_{\lambda,i}(\tau)c_{\lambda',i'}(\tau)][1-(1-\varepsilon_{a})a^+(\tau)a(\tau)]\nonumber \\
\end{eqnarray}
of Eqs. (\ref{eq:ur3}) and (\ref{eq:ur4}) (also the solutions of the operators  $a^{\dag}(t)c_{\lambda,i}(t)$ and $a^{\dag}(t)c^{\dagger}_{\lambda,i}(t)$)
in   Eq. (\ref{master-eq})
and taking   the initial conditions $\langle c^{\dagger}_{\lambda,i}(0)a(0)\rangle=\langle a(0)c_{\lambda,i}(0)\rangle=\langle a^{\dag}(0)c_{\lambda,i}(0)\rangle=\langle a^{\dag}(0)c^{\dagger}_{\lambda,i}(0)\rangle=0$
(the symbol $\langle  ... \rangle$ denotes the averaging  over the whole system of heat baths and oscillator), and assuming that
$\langle aa\rangle=\langle a^{\dagger}a^{\dagger}\rangle=\langle c^{\dagger}_{\lambda,i} c^{\dagger}_{\lambda',i'}\rangle =\langle c_{\lambda',i'} c_{\lambda,i}\rangle=0$,
$\langle  c^{\dagger}_{\lambda,i} c_{\lambda',i'}\rangle =\langle c^{\dagger}_{\lambda,i}c_{\lambda,i}\rangle=n_{\rm\lambda,i}\delta_{\lambda,\lambda'}\delta_{i,i'}$ (the heat baths consist of  independent oscillators), 
and $\langle  a^{\dagger}ac^{\dagger}_{\lambda,i} c_{\lambda,i}\rangle= n_{a} n_{\rm\lambda,i}$ (the mean-field approximation),
we obtain the  master-equation for the occupation number $n_{a}=\langle a^{\dag}a\rangle$ of the  oscillator ($a=f$  and $a=b$ for fermionic and bosonic
systems, respectively) \cite{PRE2020}:
\begin{eqnarray}
\frac{dn_{a}(t)}{dt}&=&\sum_{\lambda,i}\int^t_0ds\left\{ W^{-}_{\lambda,i}(t-s)[{\bar n}_{a}(s)n_{\rm\lambda,i}(s)-n_{a}(s){\bar n}_{\rm\lambda,i}(s)]\right.\nonumber \\
&+&\left. W^{+}_{\lambda,i}(t-s)[{\bar n}_{a}(s){\bar n}_{\rm\lambda,i}(s)-n_{a}(s)n_{\rm\lambda,i}(s)] \right\},
\label{me-a2}
\end{eqnarray}
where
\begin{eqnarray}
W^{-}_{\lambda,i}&=&\frac{2\alpha_{\lambda,i}^2}{\hbar^2}\cos([\omega-\omega_{\lambda,i}][t-s]), \nonumber \\
W^{+}_{\lambda,i}&=&\frac{2\alpha_{\lambda,i}^2}{\hbar^2}\cos([\omega+\omega_{\lambda,i}][t-s]).
\end{eqnarray}
Here,
${\bar n}_{a}(t)=1 + \varepsilon_{a} \langle a^{\dag}a \rangle$ and
${\bar n}_{\rm\lambda,i}(t)=1 + \varepsilon_{\lambda} \langle c^{\dagger}_{\lambda,i} c_{\lambda,i}\rangle$.
One can rewrite Eq. (\ref{me-a2}) as
\begin{eqnarray}
\frac{dn_{a}}{dt}&=&\int^t_0d\tau \left\{ W_+(t-\tau){\bar n}_{a}(\tau)  -  W_-(t-\tau)n_{a}(\tau)  \right\}\nonumber\\
&=&\int^t_0d\tau \left\{ W_+(t-\tau)  -  W(t-\tau)n_{a}(\tau)  \right\},
\label{na}
\end{eqnarray}
where
\begin{eqnarray}
W_+&=& \sum_{\lambda} W^{(\lambda)}_{+}\nonumber \\
&=&\sum_{\lambda,i} [W^{-}_{\lambda,i}(t-\tau)n_{\rm\lambda,i}(\tau)+W^{+}_{\lambda,i}(t-\tau){\bar n}_{\rm\lambda,i}(\tau)],\nonumber \\
W_-&=& \sum_{\lambda} W^{(\lambda)}_{-}\nonumber \\
&=& \sum_{\lambda,i} [W^{-}_{\lambda,i}(t-\tau){\bar n}_{\rm\lambda,i}(\tau)+W^{+}_{\lambda,i}(t-\tau)n_{\rm\lambda,i}(\tau)].
\label{me-na}
\end{eqnarray}
Here, $W=W_{-}-\varepsilon_{a}W_{+}$.
The coefficient $W_+$
($W_-$) defines the rate of occupation (leaving) of the state
"a" in the open quantum system. The ratio between the $W_+$ and $W_-$ characterizes the rate of equilibrium.
The occupation number reaches the equilibrium value if the ratio of $W_+$ and $W_-$ has
asymptotic at $t\to \infty$.

As shown in Refs. \cite{PhysicaA2019,PhysicaA},
for the fermionic (${ a}={ f}$) or bosonic (${a}={ b}$) oscillator (with the renormalized frequency $\Omega$)
linearly fully coupled to
$N=N_f+N_b=N_{\bar{\scriptstyle  a}}+N_{\scriptstyle  a}$
heat baths with different statistics ($N_{\bar{\scriptstyle  a}}$ Fermi and $N_{\scriptstyle  a}$ Bose baths or vice versa),
the master equation (\ref{me-a2}) or (\ref{na}) can be mapped to a simple diffusion equation
\begin{eqnarray}
\frac{dn_{\scriptstyle  a}(t)}{dt}=-2\lambda(t)n_{\scriptstyle  a}(t)+2D(t),
\label{eq:namaster2}
\end{eqnarray}
provided that
\begin{eqnarray}
W-2\varepsilon_{a}\sum_{\lambda=1}^{N_{\bar{\scriptstyle  a}}}W^{(\lambda)}_+&=&2{\dot {\lambda}}(t)-4{\lambda}(t){\lambda}(t),\nonumber \\
\sum_{\lambda=1}^{N_{\bar{\scriptstyle  a}}}W^{(\lambda)}_+ + \sum_{\lambda=N_{\bar{\scriptstyle  a}}+1}^{N}W^{(\lambda)}_+&=&2{\dot {D}}(t)-4{\lambda}(t){D}(t),
\label{ABtildeddp}
\end{eqnarray}
\begin{equation}
\label{lambda}
\lambda(t)=p\lambda_{\bar{\scriptstyle  a}}(t)+(1-p)\lambda_{\scriptstyle  a}(t) - 2\varepsilon_{\scriptstyle  a}\sum_{\lambda=1}^{N_{\bar{\scriptstyle  a}}}D_{\bar {\scriptstyle  a}_{\scriptscriptstyle\lambda}}(t)
\end{equation}
and
\begin{eqnarray}
\label{Dif}
D(t)=\sum_{\lambda=1}^{N_{\bar{\scriptstyle  a}}}D_{\bar{\scriptstyle  a}_{\scriptscriptstyle\lambda}}(t)+\sum_{\lambda=N_{\bar{\scriptstyle  a}}+1}^{N}D_{{\scriptstyle  a}_{\scriptscriptstyle\lambda}}(t).
\end{eqnarray}
  Here,
we have introduced the time-dependent friction $\lambda(t)$ and diffusion  $D(t)$ coefficients (see Appendix A).
If  $\bar{ a}={ f}$  ($\bar{ a}={ b}$)
and ${\bar{  a}}_\lambda={ f}_\lambda$
(${\bar{  a}}_\lambda={ b}_\lambda$),
then
${  a}={ b}$  (${  a}={ f}$)
and ${  a}_\lambda={ b}_\lambda$
(${  a}_\lambda={ f}_\lambda$),
respectively.
The value of $p$ is defined as  $p=\sum_{\lambda=1}^{N_{\bar{\scriptstyle  a}}}\alpha_{\scriptstyle\lambda}/\sum_{\lambda=1}^{N}\alpha_{\scriptstyle\lambda}$,  where
$\alpha_{\scriptstyle\lambda}$ is the coupling strength between the system and heat bath labeled by $\lambda$ ($\lambda=1,...,N$).
 The time-dependent friction
$\lambda_{\scriptstyle f}(t)$ [$\lambda_{\scriptstyle b}(t)$] and partial diffusion $D_{{\scriptstyle  f}_{\scriptscriptstyle\lambda}}(t)$ [$D_{{\scriptstyle  b}_{\scriptscriptstyle\lambda}}(t)$]
coefficients for the fermionic [bosonic] system coupled with $N$ fermionic [bosonic] heat baths are given in Appendix A.
 In the case of the non-Markovian dynamics, the baths affect
the system and vice versa.

Using   $D_{{\scriptstyle  f}_{\scriptscriptstyle\lambda}}(t)$,  $D_{{\scriptstyle  b}_{\scriptscriptstyle\lambda}}(t)$,
$\lambda_{\scriptstyle f}(t)$, and $\lambda_{\scriptstyle b}(t)$  from Eqs. (A10), (A11), and
the solution
\begin{eqnarray}
n_{\scriptstyle  a}(t)&=&e^{-2\int_0^td\tau \lambda(\tau)}
\left\{n_{\scriptstyle  a}(0) +  2\int_0^td\tau {D}(\tau) e^{2\int_0^\tau d\tau' \lambda(\tau')} \right\}
\label{natilde}
\end{eqnarray}
of Eq. (\ref{eq:namaster2}), one can  calculate the time-dependent  occupation number of the quantum system.


\subsection{Asymptotic occupation number}

Because the friction coefficient $\lambda_{\scriptstyle  b}(t)$ does not converge to a stationary value at $t\to+\infty$ (Fig. 1) \cite{Lac15,PRE18}, an asymptotic
stationary value of occupation number $n_{\scriptstyle  a}$ in Eq. (\ref{eq:namaster2}) can be reached if the condition
\begin{eqnarray}
\frac{1}{p}\sum_{\lambda=1}^{N_{\bar{\scriptstyle  a}}}I_{\bar{\scriptstyle  a}_{\scriptscriptstyle\lambda}}(\infty)=\frac{\frac{1}{1-p}\sum_{\lambda=N_{\bar{\scriptstyle  a}}+1}^{N}I_{{\scriptstyle  a}_{\scriptscriptstyle\lambda}}(\infty)}{1+\frac{2{\varepsilon_{\scriptstyle  a}}}{1-p}\sum_{\lambda=N_{\bar{\scriptstyle  a}}+1}^{N}I_{{\scriptstyle  a}_{\scriptscriptstyle\lambda}}(\infty)}
\label{ninftyf-f1-bN}
\end{eqnarray}
is satisfied \cite{PhysicaA2019,PRE2020}.
In other cases, the occupation number  remains oscillating at large time (Fig. 2) because the
friction $\lambda_{\scriptscriptstyle b}(t)$ and, correspondingly, diffusion coefficient
oscillate as a function of time (Fig. 1) \cite{Lac15,PRE18}.
To obtain Eq. (\ref{ninftyf-f1-bN}), the relation  $D_{{\scriptstyle  a}_{\scriptscriptstyle\lambda}}= \lambda_{\scriptstyle  a}I_{{\scriptstyle  a}_{\scriptscriptstyle\lambda}}$
is used at large time ($\Omega t\gg 1$).


The physical problem discussed here is considerably   simplified when the $N$ baths
have the same quantum nature.
Then, the asymptotic occupation
number is always stationary  (Fig. 2)  and   given by
\begin{eqnarray}
n_{\scriptstyle  a}(\infty) = \lim_{t\rightarrow \infty} \frac{D(t)}{\lambda(t)}=\lim_{t\rightarrow \infty} \frac{\sum_{\lambda=1}^{N}D_{{\scriptstyle  a}_\lambda}(t)}{\lambda_{\scriptstyle  a}(t)}=\sum_{\lambda=1}^{N}I_{{\scriptstyle  a}_{\scriptscriptstyle\lambda}}(\infty)=I_{\scriptstyle  a}(\infty)
\label{eq:asymfb0}
\end{eqnarray}
in the case when all reservoirs and system oscillator  have the same quantum nature [$p=0$, $N_{\bar{\scriptstyle  a}}=0$, $N_{\scriptstyle  a}=N$]
or
\begin{eqnarray}
n_{\scriptstyle  a}(\infty) =\frac{\sum_{\lambda=1}^{N_{\bar{\scriptstyle  a}}}I_{\bar{\scriptstyle  a}_{\scriptscriptstyle\lambda}}
(\infty)}{1-2{\varepsilon_{\scriptstyle  a}}\sum_{\lambda=1}^{N_{\bar{\scriptstyle  a}}}I_{\bar{\scriptstyle  a}_{\scriptscriptstyle\lambda}}
(\infty)} = \frac{I_{\bar{\scriptstyle  a}}(\infty)}{1-2{\varepsilon_{\scriptstyle  a}}I_{\bar{\scriptstyle  a}}(\infty)}
\label{eq:asymfb1}
\end{eqnarray}
in the case when all reservoirs have the same quantum nature (${\bar{ a}}= { b}$ or $f$)
which differs from the one of the system oscillator (${  a}= { f}$ or $b$) [$1-p=0$, $N_{\bar{\scriptstyle  a}}=N$, $N_{\scriptstyle  a}=0$].
Equations (\ref{eq:asymfb0}) and  (\ref{eq:asymfb1})   generalize
the equations given in Ref. \cite{PhysicaA} for a single bath.
If the baths have the same temperatures, then the asymptotic occupation number
differs in general from the   Fermi-Dirac or Bose-Einstein occupation number.
Only in the Markovian weak-coupling limit and in the case of the same temperature $T_\lambda=T$ of all baths,
Eqs. (\ref{eq:asymfb0}) and  (\ref{eq:asymfb1}) are reduced to the usual
Bose-Einstein and Fermi-Dirac  thermal distributions
$n_{\scriptstyle  a}(\infty)=[\exp(\hbar\omega/kT)-{\varepsilon_{\scriptstyle  a}}]^{-1}$
and the system has a thermal equilibrium.




\begin{figure}
\begin{center}
\includegraphics[scale=0.85,angle=0]{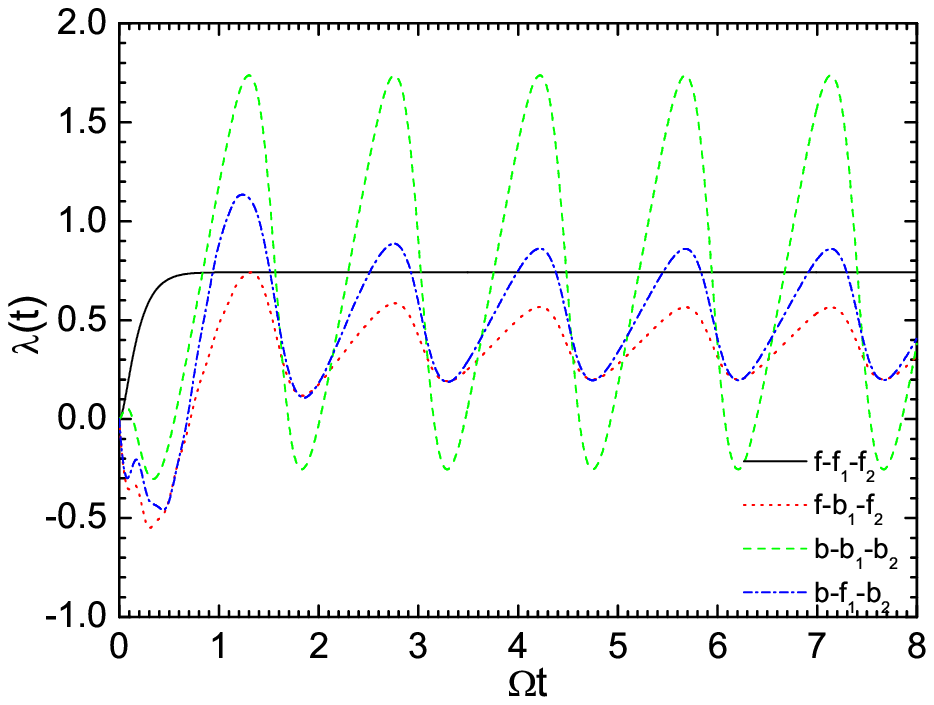}
\includegraphics[scale=0.85,angle=0]{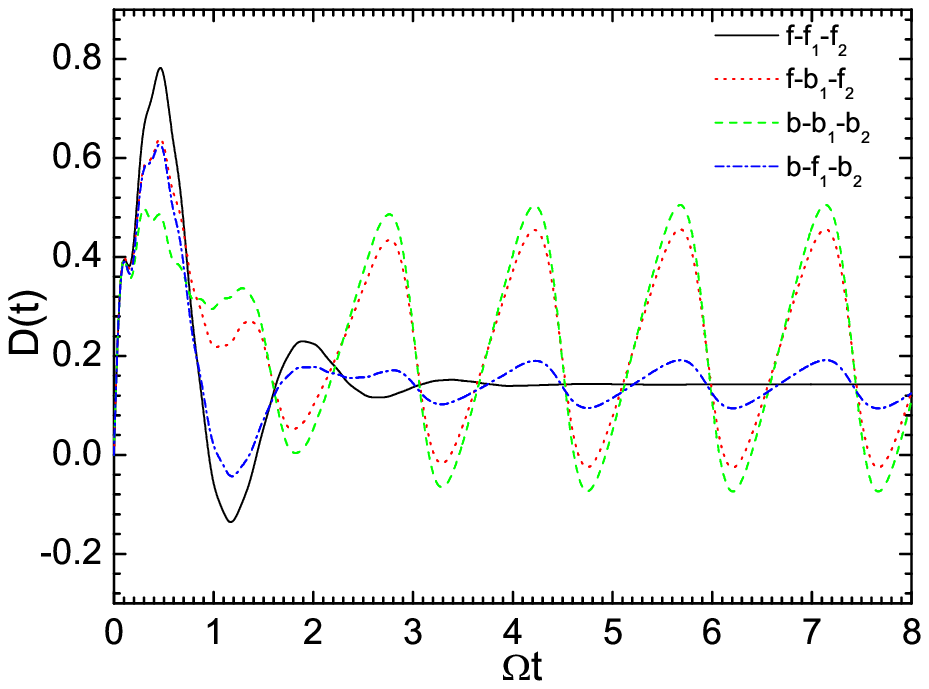}
\end{center}
\vspace{1cm}
\caption{
The calculated dependencies of the friction and diffusion coefficients on time $t$ for the
fermionic-fermionic-fermionic
($f-f_1-f_2$, solid line), bosonic-bosonic-bosonic
($b-b_1-b_2$, dashed line),
the mixed fermionic-bosonic-fermionic ($f-b_1-f_2$, dotted line),
and bosonic-fermionic-bosonic ($b-f_1-b_2$, dash-dotted line)  systems.
%
}
\label{fig:l}
\end{figure}

\begin{figure}[h]
\includegraphics[scale=0.85]{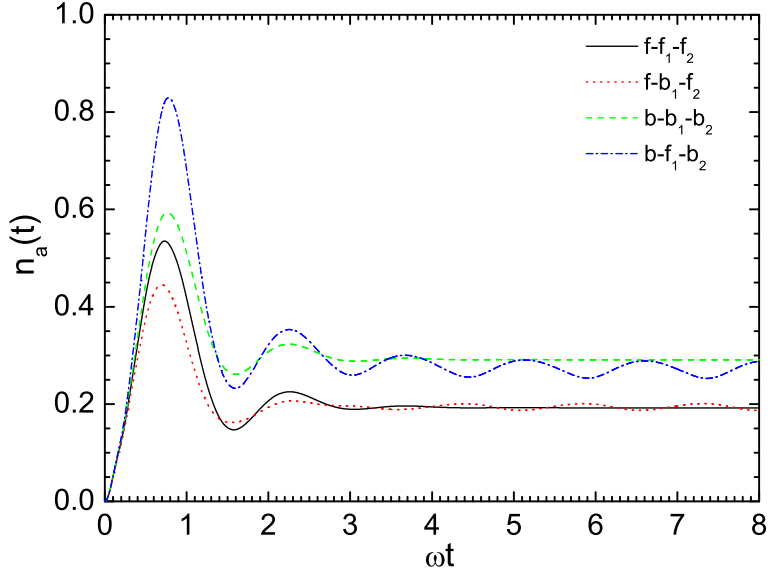}
\caption{
For the fermionic-fermionic-fermionic
($f-f_1-f_2$, solid line), bosonic-bosonic-bosonic
($b-b_1-b_2$, dashed line),
the mixed fermionic-bosonic-fermionic ($f-b_1-f_2$, dotted line),
and bosonic-fermionic-bosonic ($b-f_1-b_2$, dash-dotted line)  systems,
the calculated dependencies of the average occupation numbers on time $t$.
The plots  correspond to   initially unoccupied, $n_{\text{a}}(t=0)$=0  system state.
}
\label{fig:2}
\end{figure}

\begin{figure}[h]
\includegraphics[scale=0.85]{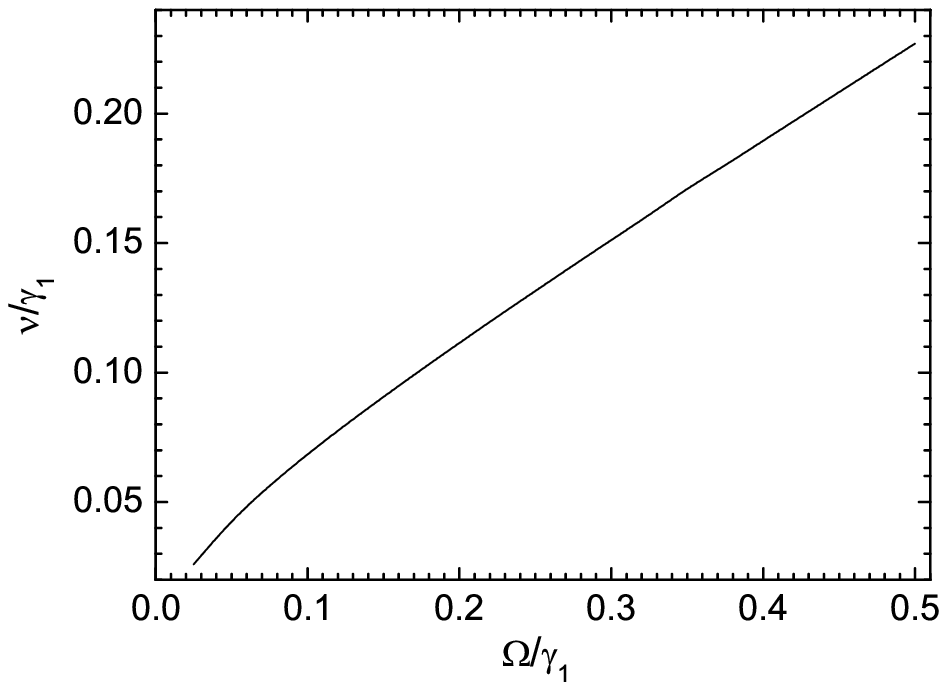}
\caption{
For
the mixed fermionic-bosonic-fermionic ($f-b_1-f_2$),
and bosonic-fermionic-bosonic ($b-f_1-b_2$)  systems,
the calculated
dependence of the frequency of oscillations of $n_{\scriptstyle  a}(t)$ at large $t$ on the oscillator renormalized frequency $\Omega$.
For two systems, the results of calculations are coincide.
}
\label{fig:3}
\end{figure}
%
%
\section{Calculated results for fermionic or bosonic oscillator coupled with  fermionic and bosonic baths
in the case of Ohmic dissipation with Lorenzian cutoffs}
\label{sec:level5}


In all figures of this paper presented for  the fermionic or bosonic oscillator  with   two baths of the same or different statistics,
we set    $\gamma_{1}/\Omega=10$, $\gamma_{2}/\Omega=15$,  $\alpha_{1}=0.1$, $\alpha_{2}=0.05$,  $g_0=\alpha_1+\alpha_2=0.15$,
$kT_1/(\hbar\Omega)=1$,  and $kT_2/(\hbar\Omega)=0.1$.
The values of $\gamma_{1,2}/\Omega$
are taken to hold the conditions $\gamma_{1,2}\gg\Omega$:
the non-Markovian quantum Langevin approach can be applied when
the system is slow in comparison  to the relaxation times of the heat baths.
The occupation numbers, diffusion and friction coefficients depend on the values of oscillator frequency
$\omega$, coupling strengths $\alpha_1$, $\alpha_2$,  inverse memory times $\gamma_{1,2}$, and heat bath temperatures
$T_{1,2}$  (see Appendix A).
A zero chemical potential is assumed here.
The values of   $\alpha_1$ and $\alpha_2$ are chosen to have the realistic values of friction coefficients
which are known from the microscopic calculations. Indeed, these coupling strengths provide almost
the same friction coefficient for relative motion of two nuclei like in Refs.~\cite{Wash1}.
As an example of bosonic system, the atomic or nuclear molecular state can be considered.
The bound or quasi-bound particle (electron in the trap or nucleon in the isomeric state) can be taken
as an example of fermionic system. The electromagnetic and temperature fields or
phonon bath can be treated as the bosonic baths. Free electrons and inclusion in the compound can act
as the fermionic baths.

For the fermionic-fermionic-fermionic
($f-f_1-f_2$), bosonic-bosonic-bosonic
($b-b_1-b_2$),
the mixed fermionic-bosonic-fermionic ($f-b_1-f_2$),
and bosonic-fermionic-bosonic ($b-f_1-b_2$)  systems,
the time-dependent   friction and diffusion coefficients   are shown in Fig. 1.
The diffusion and friction  coefficients are equal to zero at initial time.
As seen, the time dependencies of these coefficients  are not the same for the different systems.
For the $f-f_1-f_2$ system,
the friction and diffusion coefficients relatively fast reach their asymptotic values (the
transient time  for the friction  is quite short, $\Omega t\le 0.5$), whereas in the case of $b-b_1-b_2$ and
mixed $f-b_1-f_2$, $b-f_1-b_2$  systems they oscillate with the same period of oscillations.
The amplitudes of oscillations for the system   with two bosonic   baths
are larger than those for the systems with one bosonic  bath.
For the $b-b_1-b_2$ system,  the friction and diffusion coefficients oscillate in the phase and, as a result,
the  occupation number has   asymptotic limit (Fig. 2).
In the contrast, for the mixed systems  $f-b_1-f_2$ and $b-f_1-b_2$,  the  occupation number
oscillates around  certain average value
at large times, so it has no asymptotic limit. For both systems, the periods of oscillations are the same.
The  occupation number for the  fermionic oscillator  oscillates
with the larger amplitude   than  one for the bosonic   oscillator (Fig. 2).
The absolute value of oscillations mainly depend on the coupling constants.
The  times to reach the asymptotic oscillations are almost the same for  these systems.

In the case when fermionic  bath  coexists with bosonic bath, at large times
the influence of the thermostats is minimal and reversible - it takes energy from the system and gives the same amount of energy back.
As a result, the population of the excited state(s) decreases and then increases on the same level independent of the environment.
 As shown in Fig. 3,  the period of oscillations of $n_{\scriptstyle  a}(t)$ at large $t$ depends on the frequency of oscillator and, accordingly, carries
information about the system. At $\Omega/\gamma_1>0.1$,
 the frequency of asymptotic oscillations is proportional to the oscillator frequency.
Since the asymptotic oscillations of the occupation number depend on the oscillator frequency,
this gives a new opportunity to control these oscillations by changing the oscillator frequency.
For example, by this way one can control the amplification or attenuation of signal transmission.
Since the asymptotic oscillations are independent of the medium, one can  unambiguously judge
the population of the excited state of  two-level system, which, for example, is important in  quantum computers.
In this case, it is necessary to ensure a sufficient degree of
metastability of the excited states of the quantum register.
These states  must have  sufficiently a large lifetime that determines their
relaxation to the ground state due to dissipative processes.
Such a system with non-stationary asymptotics can be used as a dynamic (non-stationary) memory system because
the information about some properties of the system (population of excitation state(s) and frequency) is preserved at large times.
So, we suggest to store information by using the non-stationary memory systems.
This idea can be effective, because
such  systems will be stable under   external conditions.
%

\section{Conclusions}
In conclusion, for the   bosonic  or  fermionic  oscillator fully coupled with
the  mixed bosonic-fermionic heat baths,
the absence of equilibrium asymptotic  of occupation number was predicted.
At large times, the period of  oscillations of occupation number depends on the frequency of oscillator and, accordingly, carries
information about the system. It is an example of nonstationary (dynamic) memory storage. Each frequency
corresponds to  certain state and can lead to the control of these states for recording data in quantum computers
and increasing  channels and speeds of communication.
As shown, this behavior is also expected for other non-stationary systems (not necessarily an oscillator fully coupled with
several fermionic and bosonic heat baths) in which  the   asymptotic  friction $\lambda(t)$ and diffusion $D(t)$
coefficients periodically oscillate out of phase.

\section*{Acknowledgments}
%
%
G.G.A. and N.V.A. were supported by  Ministry of Science and Higher Education
of the Russian Federation   (Moscow, Contract No. 075-10-2020-117).
V.V.S. acknowledges the Alexander von Humboldt-Stiftung (Bonn).
D.L. thanks the CNRS for financial support through the 80Prime program.
This work was partly supported by the IN2P3(France)-JINR(Dubna) Cooperation
Programme and DFG (Bonn, Grant No. Le439/16).

\appendix

\section{Explicit expressions for friction and diffusion coefficients of fermionic (bosonic)
oscillator  with several fermionic (bosonic) heat baths}
 Let us consider the case when all $N$ heat baths and  system oscillator with the frequency $\omega$ are either all  bosonic   or all fermionic.
For these systems, the details of
the procedure for obtaining the occupation number of system
are given in Ref. \cite{PRE18}.
Here, we directly write the final expression for
the time dependence of occupation number:
\begin{equation}
\label{nt_FC}
\begin{split}
n_{{\scriptstyle  a}}(t)=n_{{\scriptstyle  a}}(0)|A(t)|^2+[1+\varepsilon_{{\scriptstyle  a}} n_{{{\scriptstyle  a}}}(0)]|B(t)|^2+I_{{\scriptstyle  a}}(t),
\end{split}
\end{equation}
where $I_{{\scriptstyle  a}}(t)=\sum_{\lambda}I_{{{\scriptstyle  a}}_{\scriptscriptstyle \lambda}}(t)$ and
\begin{eqnarray}
\label{eq:i1}
I_{{{\scriptstyle  a}}_{\scriptscriptstyle \lambda}}(t)=\frac{\alpha_{\scriptstyle\lambda}\gamma_{\scriptstyle\lambda}^2}{\pi}\int_0^{\infty}dw \frac{w}{\gamma_{\scriptstyle\lambda}^2+w^2}\left[n^{({\scriptstyle\lambda})}(w)|M(w,t)|^2+[1+\varepsilon_{\scriptstyle\lambda} n^{({\scriptstyle\lambda})}(w)] |N(w,t)|^2\right],
\end{eqnarray}
\begin{widetext}
\begin{equation}
\label{eq:dynamic}
\begin{split}
A(t)&=\frac{1}{2}\sum_{k=1}^{N_0} \xi_ke^{s_kt}
 (s_k-s_0)\\
 &\times \left\{2s_k-i[\Omega+\omega]- 2is_k\sum_{\lambda=1}^{N}\frac{\alpha_{\scriptstyle\lambda} \gamma_{\scriptstyle\lambda}}{s_k+\gamma_{\scriptstyle\lambda}}
 \right\}\prod_{\mu=1}^{N}(s_k+\gamma_{\scriptstyle\mu})\\
 &=i\sum_{\lambda=1}^{N}\alpha_{\scriptstyle\lambda} \gamma^2_{\scriptstyle\lambda}\sum_{k=1}^{N_0} \xi_ke^{s_kt}(s_k-s_0)\frac{s_k-i\omega}{s_k+i\omega}\prod_{\mu=1, \mu\neq \lambda}^{N}(s_k+\gamma_{\scriptstyle\mu}),\\
B(t)&=\frac{i}{2}\sum_{k=1}^{N_0} \xi_k e^{s_kt}(s_k-s_0)\\
&\times \left\{ \Omega-\omega+
2s_k\sum_{\lambda=1}^{N}\frac{\alpha_{\scriptstyle\lambda}\gamma_{\scriptstyle\lambda}}{s_k+\gamma_{\scriptstyle\lambda}}\right\}\prod_{\mu=1}^{N}(s_k+\gamma_{\scriptstyle\mu})\\
&=i\sum_{\lambda=1}^{N}\alpha_{\scriptstyle\lambda} \gamma^2_{\scriptstyle\lambda}\sum_{k=1}^{N_0} \xi_ke^{s_kt}(s_0-s_k)\prod_{\mu=1,\mu\neq \lambda}^{N}(s_k+\gamma_{\scriptstyle\mu}),\\
N(w,t)&=  \sum_{k=0}^{N_0} \xi_ke^{s_kt} (is_k-\omega)\prod_{\mu=1}^{N}(s_k+\gamma_{\scriptstyle\mu}),\\
M(w,t)&= -\sum_{k=0}^{N_0} \xi_ke^{s_kt} (is_k+\omega)\prod_{\mu=1}^{N}(s_k+\gamma_{\scriptstyle\mu}),
\end{split}
\end{equation}
\end{widetext}
where
\begin{equation}
\xi_k=\prod_{i=0,\\i\neq k}^{N_0}\frac{1}{s_k-s_i}
\label{eq:xik}
\end{equation}
with $s_0 =-iw$ and
the roots $s_k$, $k=1,...,N_0$,
of the    $N_0=N + 2$  order polynomial:
\begin{eqnarray}
\left[ s^2+\omega^2  - 2  \omega \sum_{\lambda=1}^{N}\frac{ \alpha_{\scriptstyle\lambda} \gamma^2_{\scriptstyle\lambda}}{s+\gamma_{\scriptstyle\lambda}}\right] \prod_{\mu=1}^{N} (s+\gamma_{\scriptstyle\mu})=0. \label{eq:polgen2}
\end{eqnarray}
Here,
\begin{eqnarray}
\Omega=\omega-2\sum_{\lambda=1}^{N}\alpha_{\scriptstyle\lambda}\gamma_{\scriptstyle\lambda}
\end{eqnarray}
is the renormalized frequency
and $\varepsilon_{\scriptstyle  \lambda} $ is equal to 1 (-1) for the bosonic (fermionic)  heat bath "$\lambda$".

In Eq. (\ref{eq:i1}),
$n^{({\scriptstyle\lambda})}(w)=(\exp[\hbar w/(kT_{{\scriptstyle\lambda}})]-\varepsilon_{\scriptstyle\lambda})^{-1}$  is
equilibrium Fermi-Dirac (Bose-Einstein)  distribution  of the fermionic (bosonic) heat bath "$\lambda$".
The $T_{{\scriptstyle\lambda}}$ is the initial thermodynamic temperature  of the corresponding heat   bath.
Here, we  introduce the spectral density $\rho_{\scriptstyle\lambda} (w)$ of the
heat-bath excitations, which allows us to replace the sum over  $i $ by integral over the frequency $w$:
$\sum_i...\to \int_{0}^{\infty}dw\rho_{\scriptstyle\lambda} (w)...$.
For all baths, we consider the following spectral function \cite{M1}:
\begin{eqnarray}
\dfrac{\alpha_{\scriptstyle{\lambda,i}}^2}{\hbar ^2 w_{\scriptstyle{\lambda,i}}}\to\dfrac{\rho_{\scriptstyle\lambda} (w) \alpha_{\scriptstyle{\lambda,w}}^2}{\hbar^2 w}=\dfrac{1}{\pi }\alpha_{\scriptstyle\lambda}\dfrac{\gamma_{{\scriptstyle\lambda}}^2}{\gamma_{{\scriptstyle\lambda}}^2+w^2},
\label{eq_222zxc}
\end{eqnarray}
where the memory time $\gamma_{{\scriptstyle\lambda}}^{-1}$ of  dissipation is inverse to the bandwidth of the heat-bath
excitations which are coupled to the collective  system. This is the Ohmic dissipation with the Lorenzian
cutoff (Drude dissipation). The relaxation time of the heat-bath should be much less than   the
characteristic collective time.
The similarity of expressions for the occupation numbers for fermionic and bosonic systems
results from the similarity of the equations of motion for creation and annihilation operators \cite{Lac15,Lac16}.

Making derivative of Eq. (\ref{nt_FC})  in $t$
and simple but tedious algebra, we
derive the following
 differential equation for the  occupation number:
\begin{equation}
\label{dnt2}
\frac{dn_{\scriptstyle a}(t)}{dt}=-2\lambda_{\scriptstyle a}(t)n_{\scriptstyle a}(t)+2D_{\scriptstyle a}(t),
\end{equation}
where
\begin{equation}
\label{lambda}
\lambda_{\scriptstyle a}(t)=-\frac{1}{2}\frac{d}{dt}\ln\left[|A(t)|+\varepsilon_{\scriptstyle a}|B(t)|^2\right]
\end{equation}
and
\begin{eqnarray}
\label{Dif}
D_{\scriptstyle a}(t)&=&\sum_{\lambda=1}^{N}D_{{\scriptstyle a}_{\scriptscriptstyle \lambda}}(t)=\lambda_{\scriptstyle a}(t)\big[|B(t)|^2+I_{\scriptstyle a}(t)\big]+\frac{1}{2}\frac{d}{dt}\big[|B(t)|^2+I_{\scriptstyle a}(t)\big], \nonumber\\
D_{{\scriptstyle a}_{\scriptscriptstyle \lambda}}(t)&=&\lambda_{\scriptstyle a}(t)\big[J_{\scriptstyle\lambda}(t)+I_{{\scriptstyle a}_{\scriptscriptstyle \lambda}}(t)\big]+\frac{1}{2}\frac{d}{dt}\big[J_{\scriptstyle\lambda}(t)+I_{{\scriptstyle a}_{\scriptscriptstyle \lambda}}(t)\big]
\end{eqnarray}
are the time-dependent friction and diffusion coefficients, respectively.
The following decomposition   $|B(t)|^2=\sum_{\lambda}J_{\scriptstyle\lambda}(t)$ is used in Eq. (\ref{Dif}). Here,
$\lambda_{\scriptstyle a}(t=0)=D_{\text{a}}(t=0)=0$.
Therefore, we have obtained the equation for  $n_{\scriptstyle a}(t)$ which is local in time. In the
case of constant transport coefficients, this equation describes the Markovian dynamics, i.e.
the evolution of $n_{\scriptstyle a}(t)$ is independent of the past. In Eq.~(\ref{dnt2}), the transport
coefficients explicitly depend on time and the non-Markovian effects are taken into consideration through this time dependence
\cite{PRE18}. The non-Markovian feature of Eq.~(\ref{dnt2}) is well seen at
$D_{\scriptstyle a}=0$. In this case, $n_{\scriptstyle a}(t) \sim \exp\left(-2\int_0^t \lambda_{\scriptstyle a}(t) dt\right)$, i.e.
the occupation number depends on the time dependence of $\lambda_{\scriptstyle a}$.
Because $A(\infty)=B(\infty)=0$ \cite{PRE18},  the appropriate asymptotic
equilibrium distribution
\begin{eqnarray}
n_{\scriptstyle a}(\infty)=\frac{D_{\scriptstyle a}(\infty)}{\lambda_{\scriptstyle a}(\infty)}=\sum_{\lambda}I_{{\scriptstyle a}_{\scriptscriptstyle \lambda}}(\infty)
\label{eq_diffasym}
\end{eqnarray}
is achieved [see Eqs.(\ref{nt_FC}) and (\ref{dnt2})]. Using Eqs.~(A1) and (A2), the asymptotic values of
$|M(w,t)|^2$ and $|N(w,t)|^2$  are found. With these values we obtain from (\ref{eq:i1})
\begin{eqnarray}
I_{{\scriptstyle a}_{\scriptscriptstyle \lambda}}(t\to\infty)&=&\frac{\alpha_{\scriptstyle\lambda}\gamma_{\scriptstyle\lambda}^2}{\pi}\int_0^{\infty}dw \frac{w}{\gamma_{\scriptstyle\lambda}^2+w^2}
\left\{[\omega+w]^2n^{({\scriptstyle\lambda})}(w)+[\omega-w]^2[1+\varepsilon_{\scriptstyle\lambda} n^{({\scriptstyle\lambda})}(w)]\right\}\nonumber\\
&\times&\frac{\prod_{\mu=1}^{N_b} (\gamma_{\scriptstyle\mu}^2+w^2)}{\prod_{k=1}^{N_0} (s_k^2+w^2)}.
\label{I19}
\end{eqnarray}
 The specific quantum nature
of the baths enters into the diffusion coefficient through the appearance of occupation probabilities.
The asymptotic diffusion and friction coefficients
are related by the well-known fluctuation-dissipation
relations connecting diffusion and damping constants.
Fulfillment of the fluctuation-dissipation relations means that we have correctly defined the dissipative
kernels in the non-Markovian  equations of motion.
In the Markovian   limit (weak-couplings and high temperatures),
the asymptotic occupation number  is:
$$n_{\scriptstyle a}(\infty)=\frac{1}{g_0}\sum_{\lambda}\alpha_{\scriptstyle\lambda} n^{({\scriptstyle\lambda})}(\omega),$$
where $g_0=\sum_{\lambda}\alpha_{\scriptstyle\lambda}$.

\end{document}